# The Myth of Global Science Collaboration

Collaboration patterns in epistemic communities.

Stefan HENNEMANN, Justus-Liebig-University Giessen, Economic Geography, Senckenbergstrasse 1, D-35390 Giessen, GERMANY (Corresponding author)

Diego RYBSKI, Potsdam Institute of Climate Impact Research PIK, P.O. Box 60 12 03, D-14412 Potsdam, GERMANY

Ingo LIEFNER, Justus-Liebig-University Giessen, Economic Geography, Senckenbergstrasse 1, D-35390 Giessen, GERMANY

Additional corresponding author information:
stefan.hennemann@geogr.uni-giessen.de
Phone: +49(0)641/99-36222
Fax: +49(0)641/99-36229

**Abstract**

Scientific collaboration is often perceived as a joint global process that involves researchers worldwide, regardless of their place of work and residence. Globalization of science, in this respect, implies that collaboration among scientists takes place along the lines of common topics and irrespective of the spatial distances between the collaborators. The networks of collaborators, termed 'epistemic communities', should thus have a space-independent structure. This paper shows that such a notion of globalized scientific collaboration is not supported by empirical data. It introduces a novel approach of analyzing distance-dependent probabilities of collaboration. The results of the analysis of six distinct scientific fields reveal that intra-country collaboration is about 10-50 times more likely to occur than international collaboration. Moreover, strong dependencies exist between collaboration activity (measured in co-authorships) and spatial distance when confined to national borders. However, the fact that distance becomes irrelevant once collaboration is taken to the international scale suggests a globalized science system that is strongly influenced by the gravity of local science clusters. The similarity of the probability functions of the six science fields analyzed suggests a universal mode of spatial governance that is independent from the mode of knowledge creation in science.

# 1 Introduction

Researchers worldwide often find themselves in a paradoxical situation: their work is being assessed on an international scale – based on international publications and citations – but it then becomes worked into national university rankings and league tables, and into local university marketing. From this paradox, three questions emerge: firstly, what is science - an international or a national endeavor? Secondly, which spatial scale has the strongest impact on science - the international, the national, or even the local scale? And thirdly, on which of these spatial scales are processes of knowledge creation organized, i.e. what is the spatial structure of the epistemic communities that are formed by the scientists of a particular field?

There are good reasons to believe that the international or global scale might be most prominent: the topics of most fields are of universal interest, the highly regarded journals are international, and the most important conferences involve participants from many countries. However, important forces act on the national scale: the large funding bodies have a clear national focus, being funded by taxpayers. Competition between universities and programs is mainly national, and this holds true to some extent for the hiring of scientists as well. One may argue that even the local scale has a strong role: local companies may sponsor and shape the directions of scientific activities, and clusters of research organizations may evolve into local centers of excellence in particular fields.

Hence, the notion that science is a borderless human endeavor may not be true when looking in detail at the spatial structures of scientific activity, and at the forces that shape these structures.

This paper aims to contribute to the understanding of one part of this puzzle: the spatial structures of epistemic communities, i.e. the spatial structures of the networks formed by collaborating scientists. This paper further focuses on investigating the following research question: to what degree does the probability of scientific collaboration depend on the physical distance between collaborators?

# 2 Theory

Some authors argue that the world is increasingly flattening out, making local peculiarities less relevant in most socially driven systems. In this discourse on the "death of distance" or the "death of the nation state" (cf. Cairncross, 1997; O'Brian, 1992; Toffler, 1980), it is not only the improving communication infrastructure as well as cheaper and faster travel that is held responsible for the increasing marginalization of geographic space. Similarities in research communities are also considered to be important proxies for space-independent collaboration. Members of an *epistemic community* may be characterized by shared notions or beliefs about subject-specific applications and techniques (cf. Haas, 1992; Weisberg & Muldoon, 2009). This should enable all members to understand one another easily. With increasing improvements in technical infrastructure, it is the *cognitive proximity* (cf. Boschma, 2005; Nooteboom, Vanhaverbeke, Duysters, Gilsing & Van den Oort, 2007) that enables members of the epistemic community to compensate for a possible lack of spatial proximity, also transcending organizational boundaries (cf. Gertler, 2003) and forming global science networks of researchers who specialize in similar fields (Moodysson, 2008).

Especially at the frontiers of science, new knowledge can be expected to be created through joint efforts in international teams of excellence (Power & Malmberg, 2008). Consequently, some empirical findings suggest a reinforcement of global structures (Wagner & Leydesdorff, 2005) and an increase in international collaboration activity in various scientific fields, such as polymers and physics, but also in theory-driven fields such as mathematics (Wagner, 2005). This trend of increasingly international research was recently confirmed by a large-scale bibliometric analysis presented by Waltman, Tijssen, and Van Eck (2011). Similarly, Moodysson, Coenen, and Asheim (2008) found that scientific collaboration in the life sciences tends to be non-local. With the advancement of certain countries, Wagner and Leydesdorff (2005) argue that the pool of potential partners will increase, since several currently peripheral countries are becoming increasingly capable, which will result in an even greater internationalization of science. However, the effective collaboration pattern is often not based simply on the "scholarly ground" of common thinking (Hoekman, Frenken & Van Oort, 2009), but is still confined within national barriers, as empirical findings of other authors suggest. For example, Almeida, Pais, and Formosinho (2009) as well as Maggioni and Uberti (2009) report strong neighboring effects in research patterns and low collaboration distances in Europe with similar overall research profiles of bordering countries. Maggioni and Uberti found that the similarity of the research patterns affects the propensity for the organizations to collaborate. These empirical findings support the theoretical notion of the need for face-to-face contacts despite the existence of cognitive proximity. It may be explained by the nature of highly unstructured and complex content in scientific knowledge creation processes (cf. Morgan, 2004, Rodríguez-Pose & Crescenzi, 2008).

This obvious ambiguity of empirical findings may be related to the scientific field under investigation. Indeed, there is consensus that subject-specific cultures affect collaboration patterns and spatial dependencies. Wagner (2005) thus suggests distinguishing between four principal scientific fields: 1) data-driven (e.g. biomedical, virology), 2) resource-driven (e.g. seismology, zoology), 3) equipment-driven (e.g. polymers, manufacturing), and 4) theory-driven (e.g. mathematics, economics). Other authors, such as Moodysson et al. (2008), suggest a distinction between analytic modes of knowledge-producing communities (i.e. natural science-oriented), and synthetic modes (i.e. more engineering-related). For an empirical assessment of the spatial patterns of epistemic communities, it is thus essential to take into account the mode of knowledge creation. As it is impossible to control for this quantitatively, empirical investigations must include different fields that encompass the different modes of knowledge creation.

In this article, the collaboration activity in scientific communities is measured from co-authorships of scientific articles. Global activities in science should therefore manifest themselves in highly international networked communities, in which all members of the network are equally likely to serve as collaborators. Unlike other approaches of measuring internationalization as an increase in mean distances over time (cf. Waltman et al., 2011), we estimate the probability of choosing a co-author from a given set of all potential collaborators in the scientific community. Therefore, the concept of epistemic communities serves as a theoretical model representing a normalized expectation of the collaboration activity in a given scientific community.

## 3 Materials and Data Organization

Table 1: Description of the raw publication data

| Technology, method, research field | Principal mode of knowledge production | Articles | | No. of individual organizations | Search string[a] |
|---|---|---|---|---|---|
| | | Raw no. | Included in analysis | | |
| Bluetooth | Synthetic, equipment | 2,171 | 274 | 455 | TS=(bluetooth) |
| Image Compression | Synthetic, theoretic | 2,399 | 403 | 614 | TS=(image compression algorithm) |
| Heart Valve | Mixed, equipment | 1,483 | 603 | 1,034 | TS=(heart valve) |
| H5N1 | Analytic, data | 1,787 | 934 | 1,271 | TS=(h5n1) |
| Tissue Engineering | Mixed, equipm./data | 8,821 | 4,240 | 2,721 | TS=(tissue engineering) |
| Carbon Nanotubes | Mixed, resource/equipm. | 29,076 | 12,656 | 4,483 | TS=(nanotube* SAME carbon) |

[a] Databases=SCI-EXPANDED, SSCI; A&HCI, CPCI-S, CPCI-SSH Timespan=2004-2008

Note: the difference between the raw no. of articles and those included in the analysis is related to articles that are written by one author or for which multiple authors share the same affiliation.

The empirical data were collected from the SCI-Expanded database of Thompson Scientific®. Bibliographic information from six distinct and sufficiently narrow scientific fields / technologies were used to capture differences in scientific communities, while common publications of organizations were considered to represent collaboration. The fields include research on the Bluetooth technology, research on image compression algorithms, heart valve research, research on the bird flu virus H5N1, tissue engineering-related research, and research that is concerned with nanotubes built from carbon. These fields encompass the different modes of knowledge creation introduced above. As a result, the individual datasets are largely internally homogeneous and mostly different from each other (see Tab 1.).

The raw publication data has been used to create undirected collaboration networks based on co-authorships between different organizations. Nodes, representing the organizations in an epistemic community (e.g. research organizations, public authorities, universities and companies), are linked to each other if authors of two organizations have at least one common publication. Only cross-organization co-authorships have been included in order to eliminate intra-organizational research team effects from the outset.

The procedure for building the empirical network for each of the six datasets consists of the following steps:

1. Select a publication with two or more individual co-authors from different organizations using the data on the affiliations of the authors.
2. Completely connect all contributing organizations of a paper with each other.
3. Repeat steps 1 and 2 for all publications in the given scientific field in the given period of time.

All organizations were geo-coded and the latitude and longitude values of all nodes were used to calculate the shortest distance $d$ along great circles on the globe.

The network representation was chosen because it provides a straightforward translation of the theoretical concepts. It also eases the calculation of randomized null models for comparison and testing, as the following section will show.

**4 Analysis**

The central idea of this approach is to compare collaboration probabilities of empirical networks with their respective randomized versions. Therefore, we considered the conditional relative frequency of nodes having an edge at distance $d$. Then, the probability was defined as the number of *identified collaborations* at a specific distance divided by the number of *possible collaborations* among the players involved at (approximately) the same distance. Hence, this is a direct transfer of the theoretical concept of *epistemic communities* to an analytical concept. In order to capture the influence of the distance, the probabilities were calculated in logarithmic bins. We would like to note that this definition differs from that used by other authors, such as Yook, Jeong, and Barabási (2002).

We thus performed the following steps:

1a  Calculate the distances $d_{i,j}$ between the two nodes of all edges (with at least one common publication).
 b  Logarithmically bin the distances: $b_{i,j} =$ int(ln $d_{i,j}$).
 c  Count the number of edges in each bin, $c_1(d)$.
2a  Calculate the distances between all pairs of nodes.
 b  Logarithmically bin these distances.
 c  Count the number of distances in each bin, $c_a(d)$
3  Calculate the relative frequency, given by $p(d) = c_1(d)/c_a(d)$.

The relative frequency $p(d)$ represents an estimate of the probability of having an edge at distance $d$. To be able to control for potential country effects, we calculated the functions for inter-country and intra-country collaboration, in addition to the full collaboration networks.

Next, an uncorrelated null model was constructed adopting a randomizing approach where the constituent property of the network, the degree distribution (the degree of a node is the number of edges it has to other nodes), is preserved, but any other feature is destroyed by shuffling (cf. Maslov & Sneppen, 2002). This includes destroying any distance relations between the nodes.

The rewiring procedure for randomization preserving the degree distribution (null model) consists of two steps:

1. Randomly pick two edges, $n_a$-$n_b$ and $m_a$-$m_b$, and swap the connections: $n_a$-$m_a$ and $n_b$-$m_b$.
2. Repeat step 2 at least as many times as there are edges in the network (we used 5 times the number of edges).

Following this, the empirical network can be compared to its randomized twin. However, in order to increase the confidence about the parameters of the null model, we repeated this procedure for 100 realizations of the rewired network. From this, we calculated the corresponding average probabilities and standard deviations as a function of the distance.

**5 Results**

The map visualization[1] of the six large-scale global science networks in Fig. 1 and 2 provides new insights into the spatial organization from the macro-perspective. One common feature of all fields analyzed is the low integration of

---

[1] The links in the maps represent a co-authorship. Geometrically, the connection is drawn as a Beziér curve that is bent allowing for the Euclidian distance between the two points in space. The shorter the distance between any two nodes, the more the curve is bent. This way of presenting spatially arranged networks allows for visualization of short distances and dense areas compared to linear connections in traditional, force-directed network visualization algorithms that are optimized to a non-overlapping presentation of edges (e. g. Fruchterman-Reingold algorithm).

Africa, Australia and South America, which, in the case of Africa, can be attributed to low overall development levels and low experience in frontier science (cf. Duque et al., 2005). Moreover, the maps show that the individual fields have different spatial patterns on the global scale. The engineering and computing science-related research topics show a higher level of activity in Asia, which fits with the comparative advantage of the regional economy. Heart valve research appears to be a matter specific to North America and Europe, whereas H5N1 research in reaction to the bird flu epidemic seems to be more of a global action, including heavy involvement of Asia (the assumed origin of the virus). Tissue engineering seems to be more spatially concentrated (spikier, thin lines) than research on carbon nanotubes (more blurred appearance of the edge cloud).

The log-log plots on the right panels in Fig. 1 and 2 show the respective conditional probabilities for each of the scientific fields. In all cases, the probability of finding a collaboration drops dramatically with the distance between the organizations. It differs by a factor of about 100 between short distances (approx. 10km) and large distances (approx 10,000km). This reveals some highly interesting commonalities between the separate scientific fields. The probability according to the null model of having a co-authored publication is independent of the distance as indicated by the green curve (squares) roughly parallel to the x-axis. All indicated probabilities above these green lines represent distances that show higher probabilities than those to be expected by chance in a globalized science community and vice versa.

The difference between collaboration that occurs inside a country (red triangle-up) and those co-authorships that cross national borders (blue triangle-down) is also very pronounced. While co-authorships in the same country show a strongly decreasing relationship – due to the typical size of the countries – of the collaboration probability and the distance between the collaborators, international collaboration is almost independent of the distance. This suggests that once the collaboration partner is sought outside one's own country, the exact location is no longer relevant. Overall, international collaboration activity is 10-50 times less relevant than national collaboration.

The apparent similarity of the curves is a striking feature of the separate fields and indicates a universality of the mechanisms governing scientific collaboration across geographic space. The individual functions of the six fields were collapsed (Malmgren, Stouffer, Campanharo & Amaral, 2009) by scaling the probabilities with

$$p^*(d) = p(d) N(N-1)/(2L) \qquad (1)$$

where N is the number of players and L is the number of ties in the corresponding field. The scaled probabilities, $p^*$, take into account the different link densities of the networks.

## (a) Bluetooth

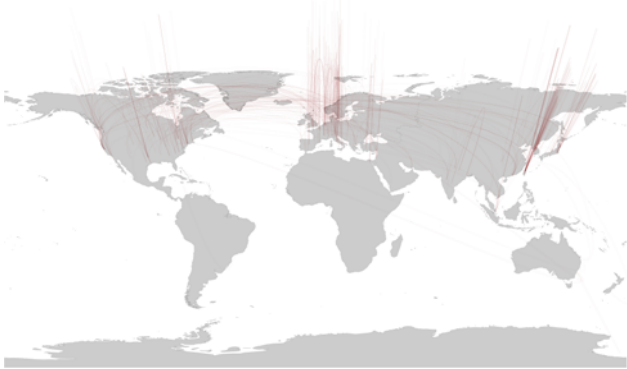
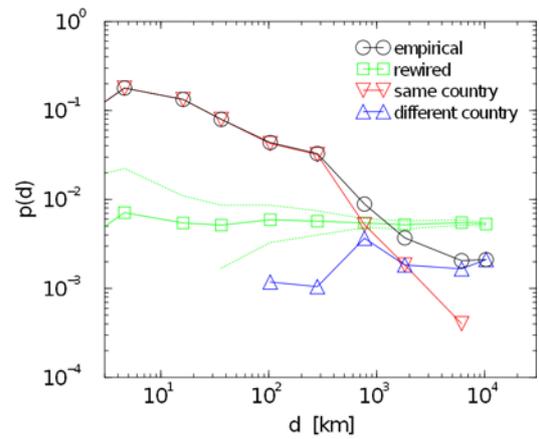

## (b) Image Compression Algorithm

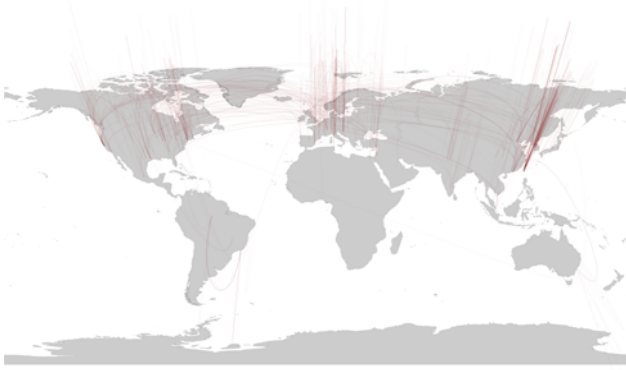
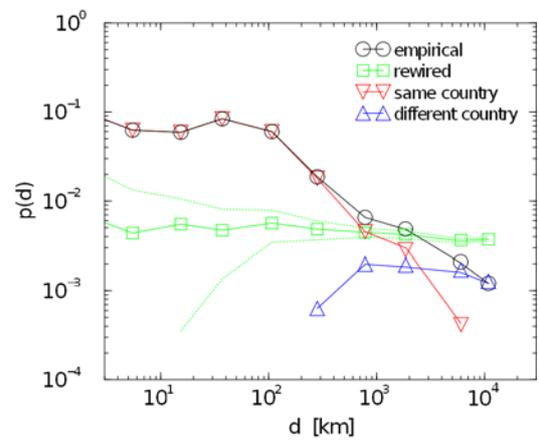

## (c) Heart Valve

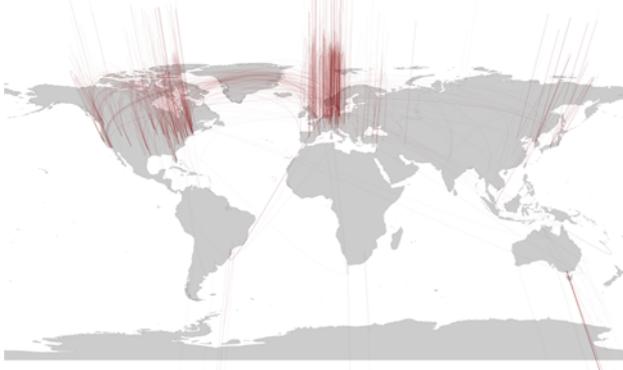
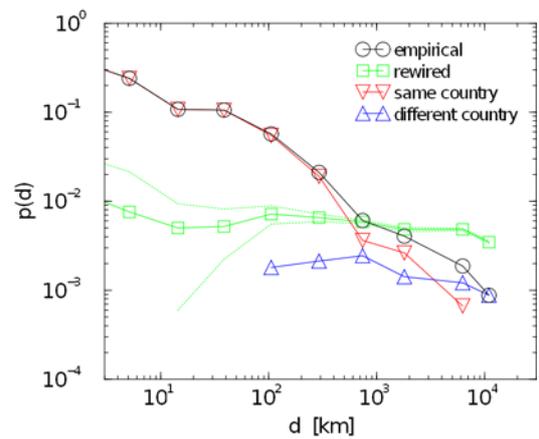

Figure 1: World maps of collaborations (left panels) and log-log plots of the conditional probability of co-authorships between organizations and the distance between them (right panels). The empirical probability of having a co-authored paper is indicated by the black circle line. The rewired random version is shown in green squares (the dotted lines represent +/-2 standard deviations of the realizations). The red triangle-down indicates the empirical probability of co-authorships within the same country, while the blue triangle-up indicates cross-country co-authorships.

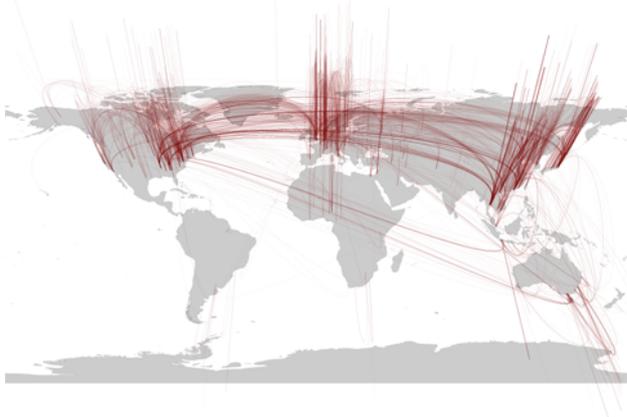 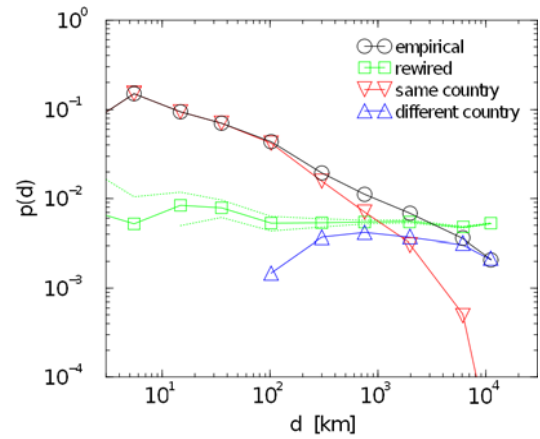

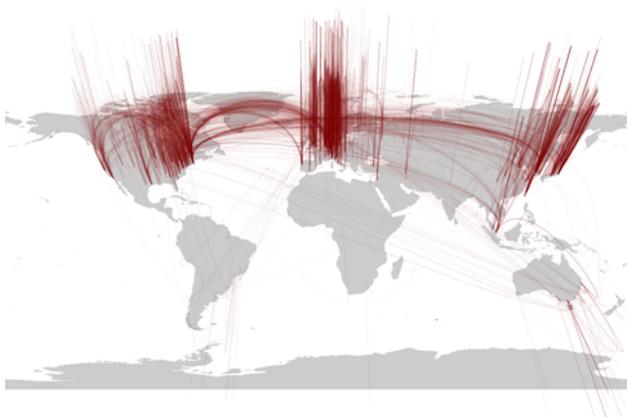 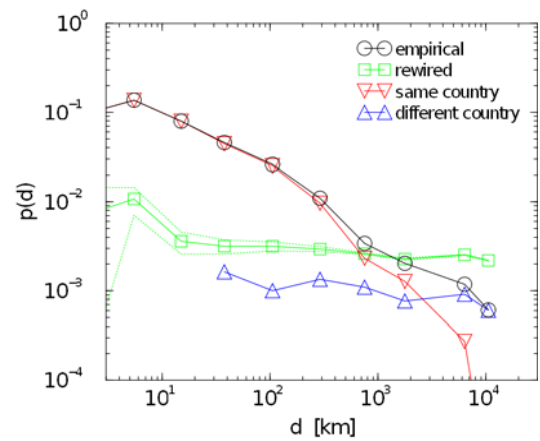

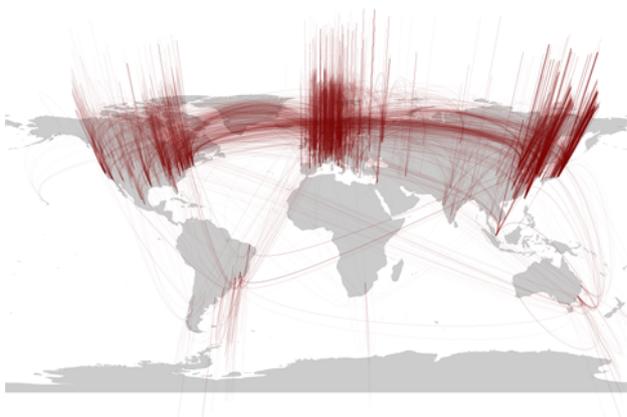 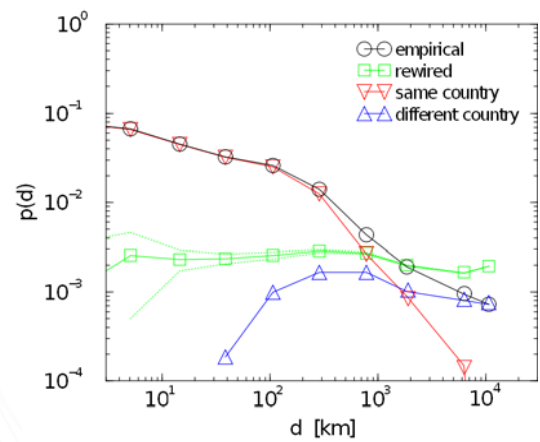

Figure 2: World maps of collaborations (left panels) and log-log plots of the conditional probability of co-authorships between organizations and the distance between them (right panels). Analogous to Fig. 1.

The resulting scaled probabilities are shown in Fig. 3. The overall scaled probability of having a co-authorship follows an almost straight power law with an exponent of -0.75 and some spreading of less than half the order of magnitude. The intra-country probabilities are very much consistent with each other, indicating a universal law of distance-dependent collaboration (Fig. 3c). Compared to this, the inter-country probability is a little more scattered, non-systematically spaced within an entire order of magnitude (Fig. 3d).

In summary, there is no sign that frontier science collaboration in epistemic communities measured as SCI-E publications is a highly internationalized activity. Moreover, national borders are a solid barrier to collaboration. This effect does not vary with the different organizational modes in individual scientific communities, with the collaboration-distance pattern in fact being uniform across scientific fields.

## 6 Discussion and conclusion

The indications for a universal mechanism are extremely surprising. In each field, we are faced with very similar spatial patterns structuring the respective epistemic community. It seems that firstly, the mode of knowledge creation is irrelevant for the spatial scope of scientific collaboration, and secondly, the spatial structure is not random. It is shaped by a few individual leading organizations in a particular field or within government initiatives etc. The spatial structures are likely to be an important property of the organization of scientific processes.

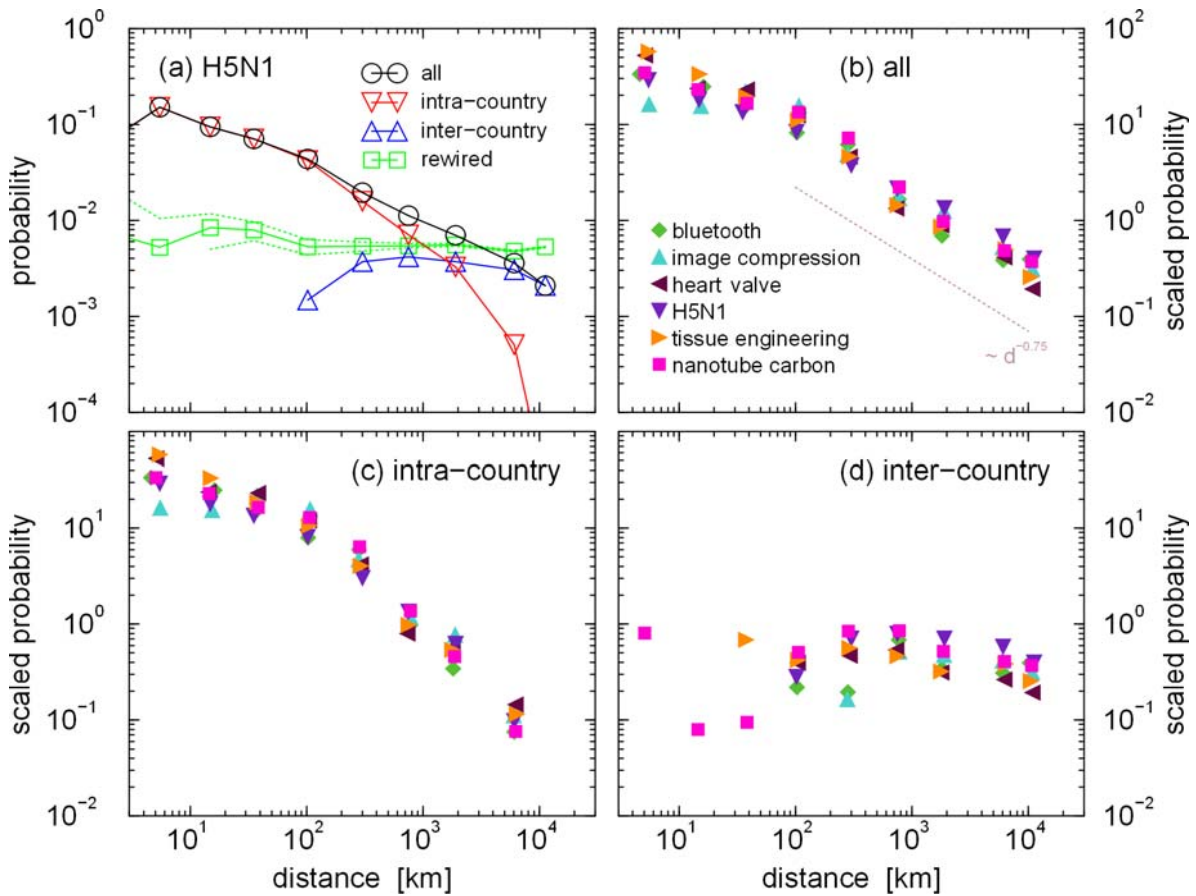

Figure 3: Probabilities of having a co-authored paper at a certain distance. The upper left plot is the same as in Fig. 2a and is just included for direct comparison purposes. The upper right plot shows all six probability functions that were scaled using Eq. 1. The lower left and right plots show the decomposed intra-/inter-country collapse of the respective six resulting curves.

Although each of the scales - global, national and local - has a role, the impact of national borders is the single most striking one. The national collaboration probability is much higher, with international collaboration thus being quantitatively of minor importance (which is not necessarily a comment on its qualitative impact, which may be tremendous). Hence, it seems that national systemic features, such as the way in which large research grants involving different organizations are set up, have the most profound impact. In addition, socio-cultural features such as language and institutions (e.g. common ethics, regulatory frameworks, legal ground or fiscal idiosyncrasies) also contribute to this nationalized pattern (cf. Müller, 2008).

Above all, distances are of great importance and place emphasis on the local scale, as the impact of the spatial distance on collaboration probability is strikingly strong. This suggests that much human interaction is involved in scientific knowledge creation and that spatial proximity is a large driver in this process. This localization behavior helps to overcome the "coordination dilemma" (Beckmann, 1993) that is inherent in situations of complex negotiation, such as research projects.

In a sense, the collaborating researchers perform a cost-benefit analysis, in which they evaluate the costs and negative externalities of maintaining the collaboration (e.g. available time, traveling costs, sharing resources, general coordination effort, unintended knowledge diffusion) against the gains (e.g. quicker publishing, future joint research proposals, access to equipment, access to complementary knowledge). However, this evaluation is usually a two-sided decision process, in which initiating a connection to a well-connected, attractive researcher has to be reciprocated (cf. Goyal, 2007). To reduce uncertainties and complexity, the attractive researchers may choose adjacent partners. The decay of $p(d)$ could then be related to the overall publication activity of the organizations. This is known to be subject to a broad distribution, i.e. a small number of organizations publish a large number of articles, while most organizations publish few articles. It is plausible that the distribution of available resources (monetary and non-monetary) follows a similar pattern. Therefore, if such resource constraints are present, the steep decay of $p(d)$ is likely to reflect the available funds (e.g. for traveling) and could be related to the publication performance of the researchers.

However, the fact that distance becomes irrelevant once collaboration is taken to the international scale saves the idea of the globalized science system. This underscores the notion that science is indeed global once it has left the strong influence of the national sphere and the gravitational pull of local science clusters. The level of globalization can be derived from the slope of $p(d)$, with flatter slopes indicating higher levels of globalized activity.

The uniformity of the spatial patterns may alternatively be explained by the fact that knowledge-producing organizations are located in cities and agglomerations, i.e. the players and ties in scientific networks are simply not randomly distributed in reality (cf. Yook et al. 2002, Rozenfeld, Rybski, Gabaix & Makse 2011, Liefner & Hennemann 2011, and references therein).

The results presented can be generalized to other scientific fields due to the robustness of the findings. This may have policy implications in the sense that the international programs and funding schemes seem to have had limited success, considering the similarity in the spatial patterns. In this context, it should also be acknowledged that global scientific progress may come not only through international collaboration, but also through heavy competition of local scientists in the arena of international science.

There are aspects of our study that need to be addressed in future work. Research indicators must not be confused with the potential quality importance of long-distance collaboration. We cannot eliminate the possibility that the few international collaborations have the greatest impact in terms of scientific progress. Related to this, it remains unclear from our results as to whether globalization of science improves the quality at all, for example if cross-country co-authorships receive more citations than purely national collaboration. At least for the organizational level, this cross-organization quality improvement has already been empirically supported (cf. Jones, Wuchty & Uzzi, 2008).

In addition, the proposed method needs to be tested with other data, such as patents or funding, to support the concept of universality in research and science-related collaboration activities.

*Conclusion*

In this article, we have proposed an approach to compare different scientific networks with one another and assess the level of globalization in these epistemic communities in research. The analytical section revealed a strongly decreasing relation between spatial distance and the probability of co-authoring a scientific publication. Moreover, this effect is much more pronounced for collaboration within countries than in cross-country collaboration. The national funding systems seem to out-compete all efforts towards a stronger international integration of scientific networks. This dominant mode of governance leads to common spatial patterns. The distances between collaborating scientists are increasing over time, as found recently by Waltman et al. (2011). However, science is far from being a globalized activity when compared with global financial markets or trade and investment flows.

The question remains: is the global competition that arises from national research organization leading towards better results on the global scale? Or would stronger global collaboration improve scientific results? To what extent does spatially clustered research interfere with research quality (which should be the focus of future research on spatial scientometrics)?